# Dynamical properties of ab initio water from machine-learning potentials

P. Montero de Hijes[1,*], L. Neubeck[1], G. Kresse[1,2], and C. Dellago[1]
[1] *Faculty of Physics, University of Vienna, A-1090 Vienna, Austria and*
[2] *VASP Software GmbH, Berggasse 21, A-1090 Vienna, Austria*

**Abstract:** We assess the dynamical properties of liquid water predicted by several density functionals using machine-learning interatomic potentials. MACE models were trained for SCAN, RPBE-D3/zd, revPBE-D3/zd, revPBE0-D3/BJ, PBE0-D3/zd, and PBE0-D3/BJ using previously reported *ab initio* datasets. We compare translational, rotational, and viscous dynamics through time-correlation functions, which resolve relaxation processes across different timescales, and through the corresponding long-time kinetic coefficients. The diffusion coefficient, second-rank orientational relaxation time, and shear viscosity reveal systematic differences among functionals. Part of these differences can be rationalized as shifts along the phase diagram, as comparisons relative to each functionals melting temperature reduce the spread in the dynamical observables. Among the functionals considered, RPBE-D3/zd provides the best overall agreement with experiment. We therefore perform a broader validation of RPBE-D3/zd using a Behler–Parrinello neural-network potential over a wide range of temperatures, densities, and pressures. The model reproduces the magnitude and anomalous pressure dependence of the diffusion coefficient, gives generally good viscosities, and captures the temperature dependence of the rotational relaxation time.

*Corresponding author:
pablo.montero.de.hijes@univie.ac.at

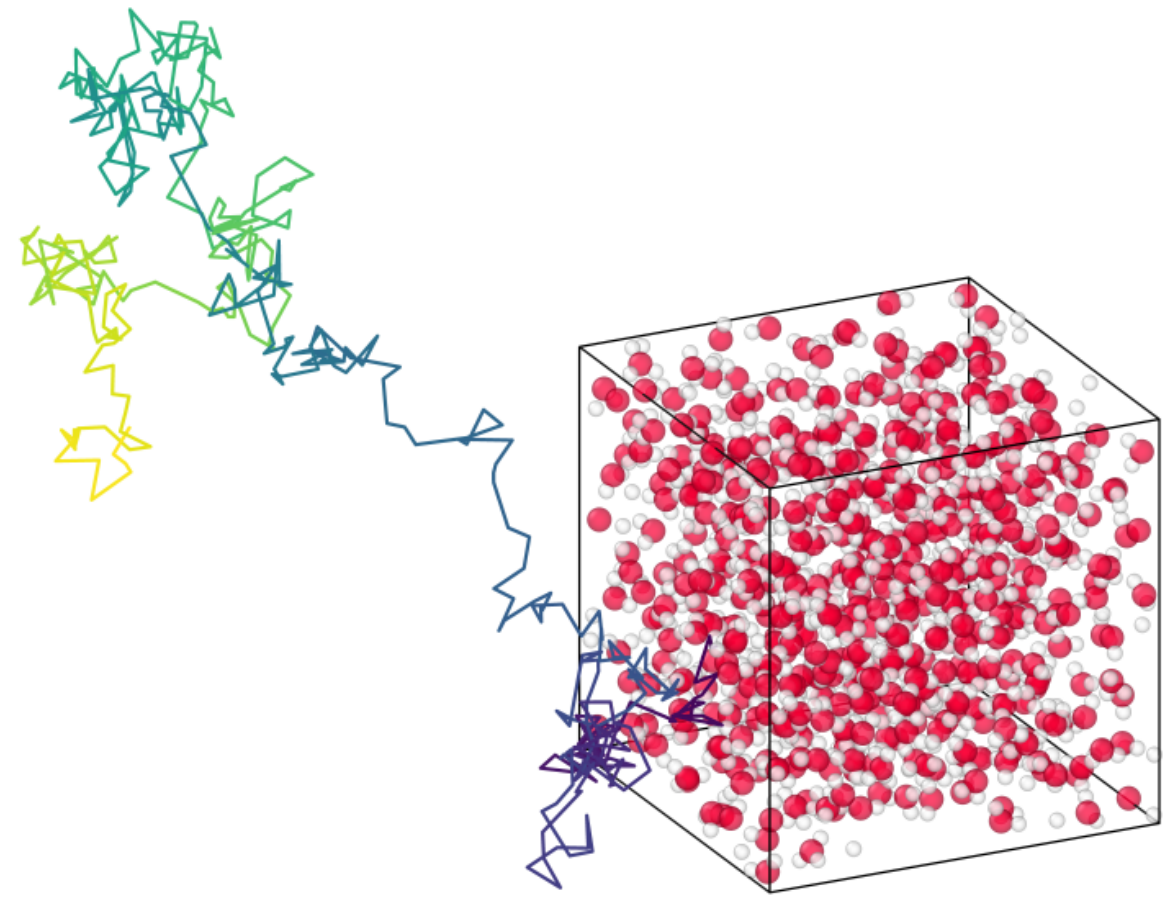

FIG. 1. Graphical abstract.

## I. INTRODUCTION

Water is one of the most studied molecular liquids, yet a quantitatively reliable first-principles description remains challenging [1]. Direct first-principles molecular dynamics simulations are limited by their computational cost, which often prevents the full convergence of dynamical and thermodynamic properties over the relevant length and time scales. Machine-learning and data-driven interatomic potentials help overcome this limitation by retaining near first-principles accuracy while enabling longer simulations, larger systems, and broader exploration of thermodynamic conditions [2–5]. In the case of water, these approaches have already shown that commonly used density functionals can yield substantially different predictions even for basic thermodynamic properties, such as the melting temperature of ice and the density isobar of the liquid [6–8].

No *ab initio* method provides a fully quantitative description of water across all relevant properties, and recent work has emphasized the importance of density-driven errors and many-body representations in DFT-based water models [5, 9]. Nevertheless, some density functionals within density functional theory (DFT) offer a reasonable compromise between computational cost and accuracy for selected thermophysical observables. However, closer agreement with experiment for certain properties, like the density, does not necessarily imply an accurate description of dynamical behavior. Transport and relaxation properties provide a particularly stringent test. Despite their importance, systematic first-principles benchmarks of water dynamics remain comparatively scarce, largely because of the high computational cost required to converge these properties [10, 11].

The dynamical characterization of water in atomistic simulations typically involves several complementary observables. Translational mobility is commonly quantified through the self-diffusion coefficient, obtained from either the mean-squared displacement or, equivalently, from the velocity autocorrelation function through a Green-Kubo relation. Rotational dynamics are characterized through orientational correlation functions and their associated relaxation times, while viscous relaxation is obtained from stress autocorrelation functions, also through Green–Kubo relations. These properties probe different, although related, aspects of molecular motion, and their comparison can reveal whether a model provides a balanced description of liquid-water dynamics [12, 13].

The main goal of this work is to assess whether the density functionals that performed best for thermodynamic properties in Ref. [6] also provide the most accurate description of water dynamics. More broadly, we aim to establish dynamical properties, in particular diffusion coefficients, rotational relaxation times, and viscosities, as complementary benchmarks for evaluating first-principles models of aqueous systems. In doing so, we also provide a set of dynamical data for several density functionals, which may be useful for future benchmarking and model development.

To this end, we first compare the dynamical properties predicted by several density functionals using accurate MACE machine-learning potentials [14]. We then focus on RPBE-D3/zd, which provides the best overall kinetic description among the functionals considered, and extend the comparison over a broader range of temperatures and pressures using the more affordable Behler–Parrinello neural-network potential [15]. The resulting diffusion coefficients, viscosities, and rotational relaxation times are compared with experimental data over a broad range of thermodynamic conditions [16–20]. Our results show that agreement with thermodynamic properties alone does not guarantee an accurate description of water dynamics. Among the functionals studied here, RPBE-D3/zd provides the most balanced description of translational diffusion, rotational relaxation, and shear viscosity, while part of the spread among functionals can be rationalized as a shift along the phase diagram relative to their respective melting temperatures.

## II. METHODS

We employ the molecular dynamics simulation package LAMMPS[21] to compare the dynamical behavior of water as predicted by different DFT functionals at ambient pressure. To this end we train MACE models for SCAN, RPBE-D3/zd, revPBE-D3/zd, revPBE0-D3/BJ, PBE0-D3/zd, and PBE0-D3/BJ using the datasets from Ref. [6]. For a more extensive comparison with experiments over a broader range of temperatures and pressures, we use the computationally less demanding Behler–Parrinello neural network potential (BPNN) reported in Ref. [3], which was trained on the RPBE-D3/zd functional.

### A. MACE models

For each density functional, we trained an independent MACE potential [14] from scratch using the corresponding DFT water dataset following the same training workflow. This means that for each functional, the dataset was split into training, validation, and test sets using an 80/10/10 split. To keep the subsets representative, configurations were grouped according to their mean force norm and then selected using Kennard-Stone sampling [22] in SOAP descriptor space [23].

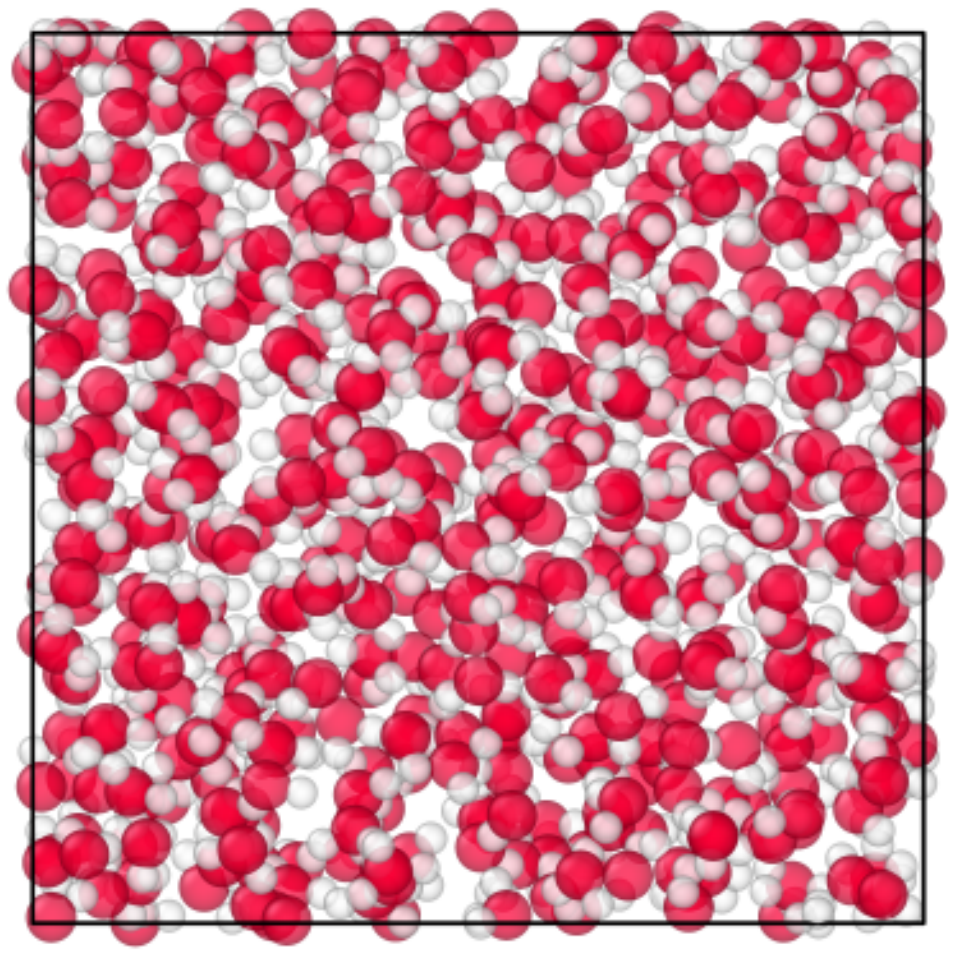

FIG. 2. Screenshot of a liquid configuration from an NVE trajectory using a MACE model trained on the SCAN functional.

For the MACE architecture, we used 64 channels, two interaction layers, maximum angular order $L = 1$, correlation order 3, and a cutoff radius of 5.0 Å. The same architecture was used for all functionals to keep the comparison consistent.

Training was performed in single precision with the Adam optimizer [24] and AMSGrad [25]. The training batch size was 8, and the validation batch size was 16. The loss combined energies and forces, with weights of 1 and 1000, respectively. Each model was trained for up to 2000 epochs. The last part of the training used the Stage Two/SWA procedure, starting at epoch 1500. In this stage, following the standard MACE training recipe, the energy contribution to the loss is increased, which helps reduce the final energy error. An exponential moving average of the model parameters was also used with a decay factor of 0.99. The final Stage Two models were exported to TorchScript/LAMMPS format and used for the molecular dynamics simulations. All model training runs were carried out on the Leonardo supercomputer at CINECA using one NVIDIA A100 GPU per training run. The validation root mean squared error (RMSE) values are reported in Tab. I.

### B. Dynamical properties

We characterize the dynamics at two complementary levels: first, through time-correlation functions, which resolve relaxation processes across different timescales, and second, through the corresponding long-time transport coefficients, which quantify diffusive behavior [26, 27]. The former approach requires a high output frequency of configurations to capture fast relaxation processes, whereas the latter emphasizes long simulation times, complementary testing the stability of the MACE models [28].

We investigate three types of dynamical properties: translational, rotational, and viscous relaxation. To characterize translational motion, we compute the velocity autocorrelation function (VACF),

$$C_{VV}(t) = \langle \mathbf{v}_{\mathrm{O}}(t') \cdot \mathbf{v}_{\mathrm{O}}(t'+t) \rangle , \tag{1}$$

where $\mathbf{v}_{\mathrm{O}}$ is the velocity of the oxygen atom in a water molecule, used here as an approximation to the molecular translational velocity, $t'$ is a time origin, and $t$ is the lag time from this time origin. The self-diffusion coefficient $D$ is then obtained through the Green–Kubo expression

$$D = \frac{1}{3} \int_0^\infty C_{VV}(t)\, dt. \tag{2}$$

For completeness, we also obtain $D$ from the translational mean-squared displacement (MSD), which is defined as

$$\left\langle \Delta r^2(t) \right\rangle = \frac{1}{N} \sum_{i=1}^{N} \left\langle \left| \mathbf{r}_i(t'+t) - \mathbf{r}_i(t') \right|^2 \right\rangle , \tag{3}$$

where the translational diffusion coefficient then follows from the long-time diffusive regime as

$$D = \lim_{t\to\infty} \frac{1}{6t} \left\langle \Delta r^2(t) \right\rangle . \tag{4}$$

To characterize rotational dynamics, we compute the second-order orientational correlation function $C_{\mathrm{OH}}(t)$, whose associated relaxation time can be obtained experimentally [18, 29]. In particular, $C_{\mathrm{OH}}(t)$ is directly related to the anisotropy decay measured in polarization-resolved femtosecond infrared pump–probe experiments on HDO in $H_2O$ [19, 30]. To track rotations, we define a unit vector describing the orientation of a water molecule. Denoting this vector by $\hat{\mathbf{u}}(t)$, we select the O–H bond vector (from the oxygen atom to a hydrogen atom), as this choice enables a more direct comparison with experimental measurements that typically employ HDO diluted in water [18, 29]. This autocorrelation function is defined as:

$$C_{\mathrm{OH}}(t) = \langle P_2(\hat{\mathbf{u}}(t') \cdot \hat{\mathbf{u}}(t'+t)) \rangle , \tag{5}$$

where

$$P_2(x) = \frac{1}{2}(3x^2 - 1) \tag{6}$$

is the second Legendre polynomial.
The second-rank orientational relaxation time was obtained by integrating the corresponding correlation function,

$$\tau_2 = \int_0^\infty C_{\mathrm{OH}}(t)\, dt. \tag{7}$$

Finally, viscous relaxation is characterized through the autocorrelation functions of $m$ independent shear-stress tensor components,

$$C_\eta(t) = \frac{1}{m} \sum_\alpha \langle p_\alpha(t') p_\alpha(t'+t) \rangle , \tag{8}$$

where here we use $m = 5$ independent traceless stress components:

$$p_\alpha \in \left\{ p_{xy},\, p_{xz},\, p_{yz},\, \frac{1}{2}(p_{xx} - p_{yy}),\, \frac{1}{2}(p_{yy} - p_{zz}) \right\}. \tag{9}$$

The shear viscosity is then obtained from the Green–Kubo relation

$$\eta = \frac{V}{k_{\mathrm{B}} T} \int_0^\infty C_\eta(t)\, dt. \tag{10}$$

We note that each configuration along the trajectory contributes $m$ values to $C_\eta$. This is significantly fewer than for the VACF, where each configuration contributes as many values as there are molecules in the system (typically two orders of magnitude larger than $m$), and in the rotational autocorrelation function, each molecule contributes twice because of the two intramolecular OH bonds. This reduced statistical sampling, combined with the often slow decay of $C_\eta$, makes the convergence of $\eta$ significantly more demanding.

Finally, we note that it is well known that $D$ is affected by finite-size effects under periodic boundary conditions. One way to correct for them is to simulate at least three sufficiently different system sizes and extrapolate to the infinite-size limit. Another approach is to correct the computed values using [31]

$$D = D_{\mathrm{PBC}} + \frac{k_{\mathrm{B}} T \xi}{6\pi\eta L}, \tag{11}$$

where $\xi = 2.837297$, $L$ is the box length, $T$ is the temperature, and $k_{\mathrm{B}}$ is the Boltzmann constant. In both cases, this implies a significant amount of computational cost, either because more simulations are required or because they have to be sufficiently large to converge the integration of $C_\eta(t)$.

### C. Molecular dynamics simulations

For the MACE models, we found that simulations containing 128 and 512 molecules had similar computational costs, since the smaller system appeared to underutilize the available computational resources. We therefore used the larger system size. For 512 molecules, finite-size effects are still expected to lead to an underestimation of $D$ by approximately 15% [3, 32, 33].

For the comparison of density functionals using the LAMMPS/MACE implementation, we employ a 0.5 fs timestep and perform two sets of simulations. The first set targets the high output frequency required to compute time-correlation functions and each simulation is carried out in the NVT ensemble for 600 ps. These simulations are used to compute translational, rotational, and viscous dynamical properties. The diffusion coefficients obtained from the VACF are corrected for finite-size effects using Eq. (11), with $\eta$ taken from the same set of simulations.The second set of simulations targets long-time diffusive behavior and each simulation is carried out in the NVE ensemble running for 5–9 ns. These NVE simulations are used to compute translational diffusion coefficients from the mean-squared displacement.

The temperatures selected for the NVE runs are chosen based on the corresponding density isobars [6], in order to sample state points around the temperature of maximum density. The temperatures of melting and maximum density from Ref. [6] are reported in Tab. I. For the NVT simulations, we select two specific temperatures: the melting temperature of the corresponding functional and 297 K, which corresponds to the average of the melting temperatures considered here, and is also a reasonable reference for ambient temperature. This choice allows us to compare the models both at the same absolute temperature and at a comparable level of thermodynamic stability with respect to ice.

To perform these simulations around ambient pressure, before starting the production runs, we first equilibrate each system for 1 ns in the NpT ensemble at a pressure of 1 bar and the target temperature $T$. We then restart the simulation in the NVT and NVE ensembles, respectively, from a configuration whose instantaneous observables are representative of the desired values along the NpT trajectory.

For the comparison between RPBE-D3/zd and experiment, we employ n2p2/LAMMPS [34] and simulate systems of 128 molecules in the NVT ensemble. We consider four isotherms, 250, 270, 290, and 310 K, spanning mildly negative to relatively high pressures, corresponding to densities of 800, 900, 1000, and 1100 kg/m$^3$. The timestep is the same as for the MACE models. Configurations are written every 0.5 ps, and $D$ is computed from the MSD of the oxygen atoms in water molecules once the diffusive regime has been reached. The second-order rotational relaxation time $\tau_2$, obtained from Eq. (7), requires a finer output frequency; therefore, we analyze trajectories every 0.05 ps. To compute $\eta$ from Eq. (10), we use the timestep, 0.0005 ps, as the smallest time lag, to accurately capture the early-time decay of the stress autocorrelation function. The final value of $\eta$ is estimated from the plateau region of the running integral, selected by requiring both a small relative variation and a negligible residual slope over the chosen time window. The reported diffusion coefficients are corrected for finite-size effects using Eq. (11).

TABLE I. Temperatures associated with each functional, including the melting temperature $T_{\mathrm{m}}$ and the temperature of maximum density $T_{\mathrm{d}}$ from Ref. [6]. The table also reports the model training accuracy in terms of root-mean-squared errors (RMSE) for energy and forces. Units are K for temperature, meV/atom for $E_{\mathrm{RMSE}}$, and meV/Å for $F_{\mathrm{RMSE}}$.

| Functional | $T_{\mathrm{m}}^{[2]}$ | $T_{\mathrm{d}}^{[2]}$ | $E_{\mathrm{RMSE}}$ | $F_{\mathrm{RMSE}}$ |
|---|---|---|---|---|
| SCAN | 299 | 321 | 0.1 | 10.3 |
| RPBE-D3/zd | 278 | 283 | 0.1 | 5.6 |
| revPBE-D3/zd | 283 | 284 | 0.1 | 5.5 |
| revPBE0-D3/BJ | 289 | 280 | 0.1 | 6.1 |
| PBE0-D3/zd | 303 | 300 | 0.1 | 6.3 |
| PBE0-D3/BJ | 330 | 327 | 0.1 | 6.2 |

## III. RESULTS

### A. Comparing density functionals

First, we compare the VACF for the different functionals, shown in Fig. 3, at two representative temperatures. The top panel corresponds to 297 K. The bottom panel shows the VACF evaluated at the relative melting point $T_m$ of each functional, i.e., $T = T_{\mathrm{m}}$. This representation may be useful for highlighting whether discrepancies observed at fixed $T$ instead reflect a systematic shift of the phase diagram, while the underlying dynamical behavior remains otherwise similar.

As shown in Fig. 3a), at 297 K, all functionals require approximately the same time to decorrelate fully. However, during this correlated period, the magnitude and timing of the correlations differ among functionals: for some, the correlations are more pronounced, while for others, they appear earlier or later. This behavior may reflect the different thermodynamic meaning of 297 K for each functional. For instance, the melting temperature is 330 K for PBE0-D3/BJ, whereas it is 278 K for RPBE-D3/zd. In panel b), where the comparison is made at the respective melting temperature of each functional, these differences are notably reduced. Nevertheless, differences remain in the shape of the VACF at short and intermediate times. These differences mainly reflect variations in the local inertial and cage dynamics rather than a markedly different overall decorrelation time, since all curves decay to zero on a broadly similar timescale.

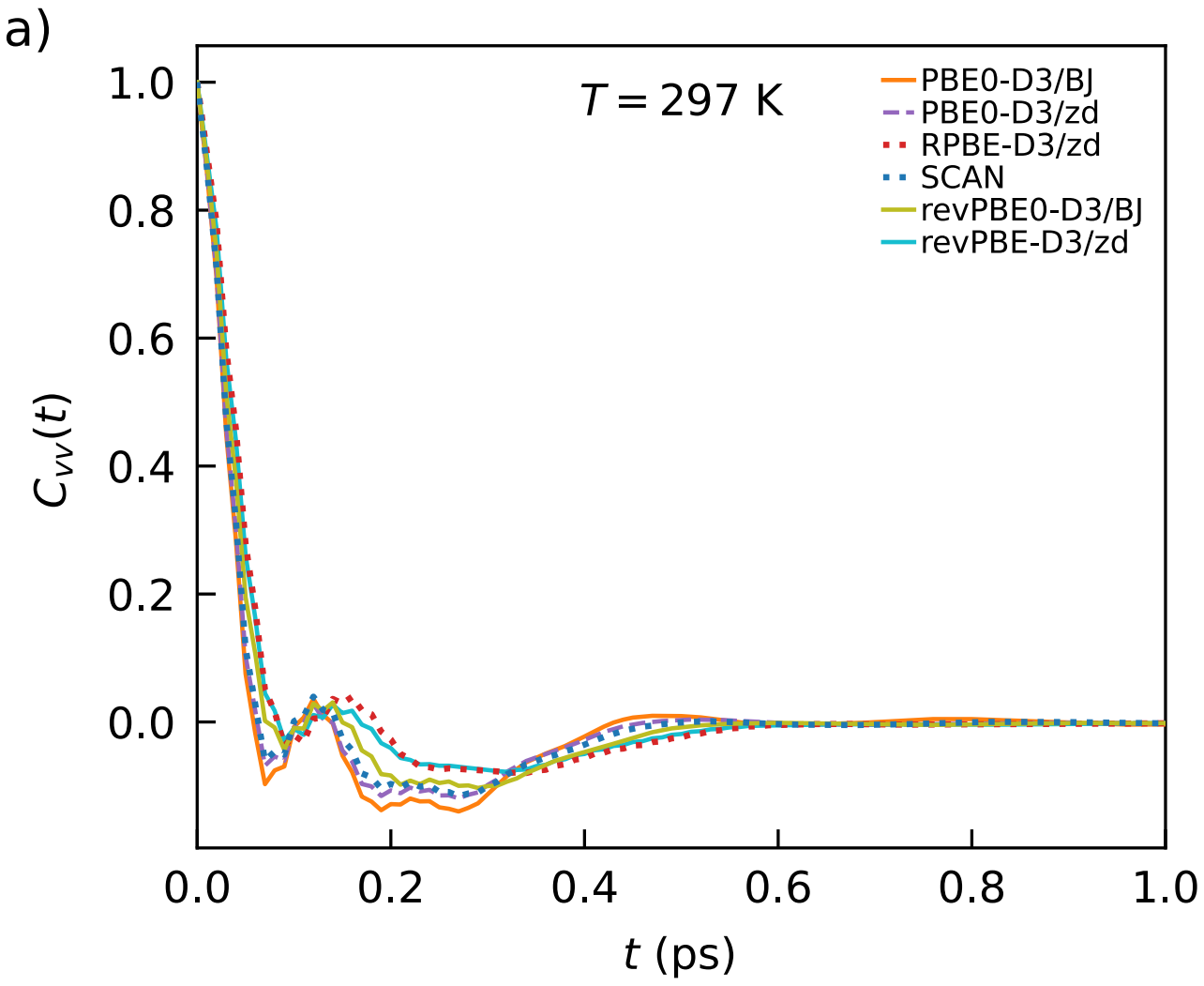

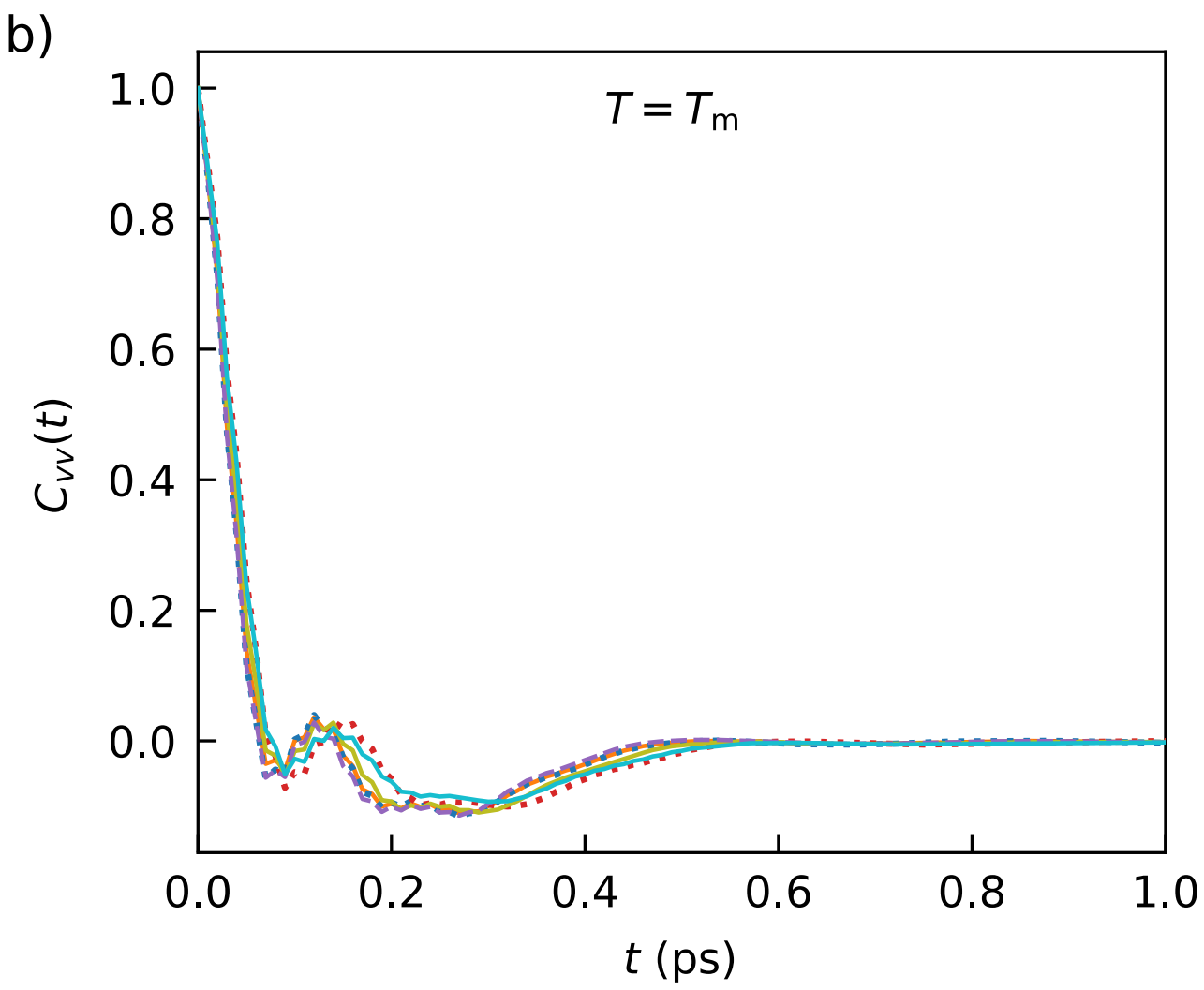

FIG. 3. Normalized velocity autocorrelation functions, $C_{VV}(t)$, of liquid water obtained from *ab initio* machine learning molecular dynamics simulations using different functionals. Panel a) corresponds to $T = 297\,\mathrm{K}$, while panel b) shows results at the respective melting temperature $T_m$ of each functional, which are given in Tab. I.

A way to quantify the differences in the shape of the VACF is to compute the diffusion coefficient $D$. We first use Eq. (2), which provides a reliable estimate from the VACF. These simulations were limited to 600 ps and employed a high output frequency to accurately resolve the short-time decay of the correlation function, which strongly influences the integral. To access longer timescales, additional simulations were performed with reduced output frequency, targeting the diffusive regime directly. These trajectories span 5–9 ns, and $D$ is obtained from the long-time behavior of the MSD.

These results are shown in Fig. 4. Panel a) compares the diffusion coefficients at fixed temperatures, whereas panel b) reports them as a function of $\Delta T = T - T_{\mathrm{m}}$, i.e., relative to the melting temperature of each model at ambient pressure. As shown in panel a), several functionals exhibit significant deviations from the experimental values. However, panel b) indicates that part of this discrepancy may be viewed as arising from a systematic shift of the phase diagram: the agreement generally improves when the temperatures are referenced to the melting temperature $T_{\mathrm{m}}$. Notably, SCAN substantially underestimates $D$ even after this rescaling. PBE0-D3/zd also shows clear deviations, while PBE0-D3/BJ appears strongly shifted relative to real water. In contrast, RPBE-D3/zd and revPBE-D3/zd yield good agreement with experiment. Finally, revPBE0-D3/BJ shows intermediate performance, being neither particularly accurate nor strongly deficient. The good agreement of RPBE-D3/zd with experiment, together with the underestimation of $D$ by SCAN, was also reported in Ref. [35]. In that work, neuroevolution potentials were trained to reproduce these density functionals, as well as the data-driven many-body potential MB-pol, which yields very good agreement with experimental diffusion coefficients.

Our estimates of $D$ differ slightly depending on whether they are obtained from the MSD in the long NVE simulations (shown as circles) or from the VACF in the NVT simulations (shown as crosses). The former values are expected to underestimate the infinite-system diffusion coefficients by approximately 15%. Since $\eta$ was not computed in this case, Eq. (11) could not be applied to correct for finite-size effects. Diffusion coefficients obtained from the VACF are corrected using Eq. (11), which accounts for part of the discrepancy between the two estimates. Additional contributions to the discrepancy may arise from the weaker control over the thermodynamic state in the NVE ensemble. The selected initial configurations and velocities led, in some cases, to pressures that deviated from ambient pressure; for example, the deviation was about $-300$ bar for PBE0-D3/zd. In all cases, however, the pressure deviation remained below approximately $\pm 350$ bar. Setting the thermodynamic state more accurately in the NVE ensemble would have substantially increased the already high computational cost of these calculations. These pressure deviations do not affect the conclusions drawn from the comparison among functionals.

Although expressing the results as a function of $\Delta T = T - T_{\mathrm{m}}$ reduces part of the discrepancy in the diffusion coefficient, this rescaling does not provide a universal correction. In particular, Ref. [6] showed that the density isobars of different functionals exhibit significant spread, which is not eliminated by a simple temperature shift. Therefore, while rescaling by the melting temperature can partially account for differences in the location of the liquid state in the phase diagram, it does not resolve the underlying discrepancies in the free energy landscape.

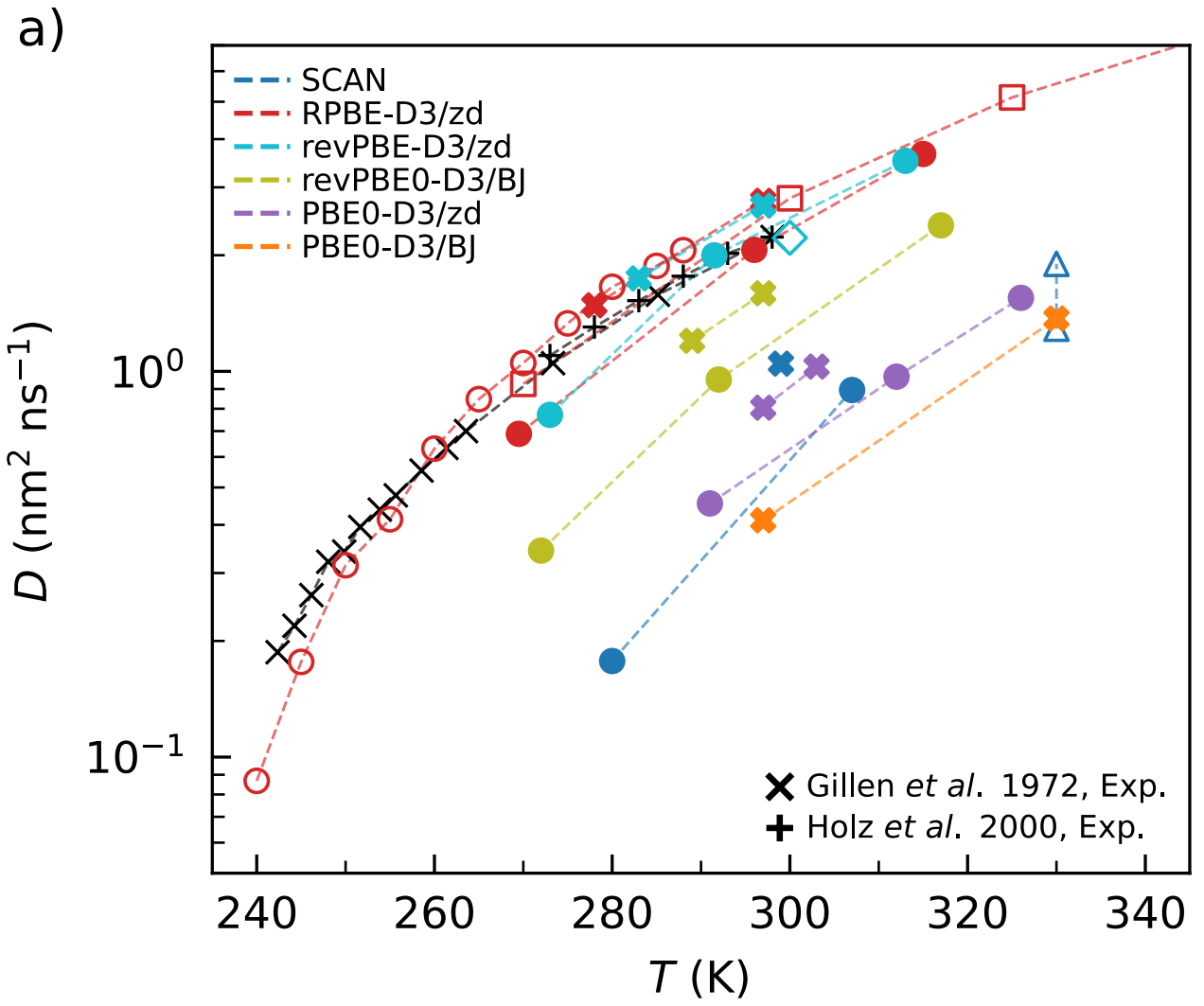

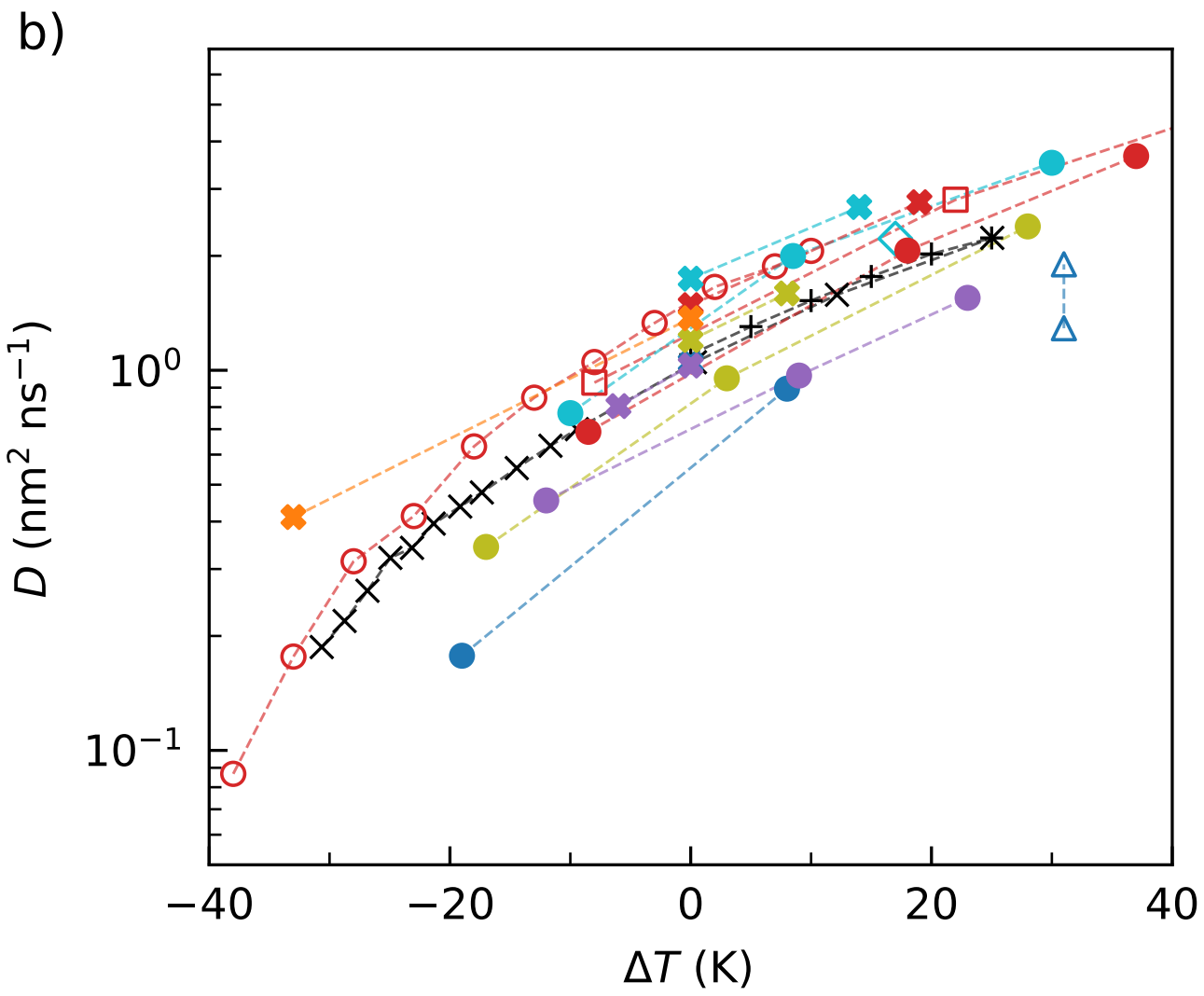

FIG. 4. a) Self-diffusion coefficient $D$ of water as a function of temperature $T$ from experiments [36, 37] and simulations based on different density functionals. Experimental data are shown in black. Colors identify the underlying density functional, as indicated in the legend. The present results, obtained using MACE models trained on the datasets of Ref. [6], are shown with filled symbols; circles denote values obtained from the MSD over a few ns of NVE simulations, whereas crosses indicate values computed from VACF calculations using 600 ps-long NVT simulations and corrected for periodic boundary effects using Eq. (11). Simulation data from the literature are shown with open symbols: RPBE-D3/zd data from a Behler–Parrinello neural network [38] and kernel regression [39], ab initio molecular dynamics data for revPBE-D3/zd [40], and ab initio molecular dynamics data for SCAN [41]. Dashed lines are used only as guides to the eye and connect points belonging to the same dataset. b) Same data as in panel a), plotted as a function of $\Delta T = T - T_\mathrm{m}$.

Next, we analyze rotational kinetics. Figure 5 shows the orientational correlation functions for the different functionals. Consistent with the translational dynamics, PBE0-D3/zd exhibits markedly slower relaxation times, reflecting a significant shift of its phase diagram. However, as shown in panel b), at the melting point, $\Delta T = T - T_\mathrm{m}$, its behavior becomes comparable to that of the other functionals. Overall, the rotational dynamics follow trends similar to the translational ones. In particular, functionals that yield comparable diffusion coefficients $D$ (such as SCAN and PBE0-D3/zd) also exhibit similar behavior in $C_\mathrm{OH}(t)$.

To enable direct comparison with experiments, we compute the second-rank relaxation time $\tau_2$ using Eq. (7), for which experimental data are available. Figure 6 presents $\tau_2$ for the different functionals, compared with experimental values. The conclusions are consistent with those drawn from translational diffusion. While most discrepancies are reduced when accounting for shifts of the phase diagram, some differences remain. In particular, out of the two best-performing functionals so far, revPBE-D3/zd shows slightly worse agreement with experiment when temperatures are rescaled relative to $T_\mathrm{m}$, so that RPBE-D3/zd provides the best agreement with experimental rotational and translational dynamics among the functionals considered.

The final comparison among functionals concerns the shear viscosity. In Fig. 7, we observe trends similar to those found for the other dynamical properties. RPBE-D3/zd and revPBE-D3/zd show the best agreement with the experiments. PBE0-D3/BJ approaches the experimental behavior only after a substantial shift along the phase diagram, as observed in panel b), whereas SCAN and PBE0-D3/zd predict dynamics that are clearly too slow compared with real water, in this case represented by larger viscosity. Finally, revPBE0-D3/BJ shows intermediate performance again and appears to be the most accurate among the hybrid functionals considered here.

### B. RPBE-D3/zd vs. Experiments

Since the comparison across functionals identified RPBE-D3/zd as providing the best agreement with experiments, we perform a more extensive evaluation of its transport properties, covering $D$, $\eta$, and $\tau_2$ based on the BPNNP. While the previous analysis was restricted to ambient pressure, we now consider a broader thermodynamic range by simulating water at densities of 800, 900, 1000, and 1100 kg/m$^3$ and temperatures of 250, 270, 290, and 310 K. These state points span pressures from mildly negative values to several hundred MPa. In the comparison with the experimental data, the experimental temperatures may differ slightly from the simulated isotherms, typically by less than 3 K. Larger differences, when present, are indicated explicitly in the legend.

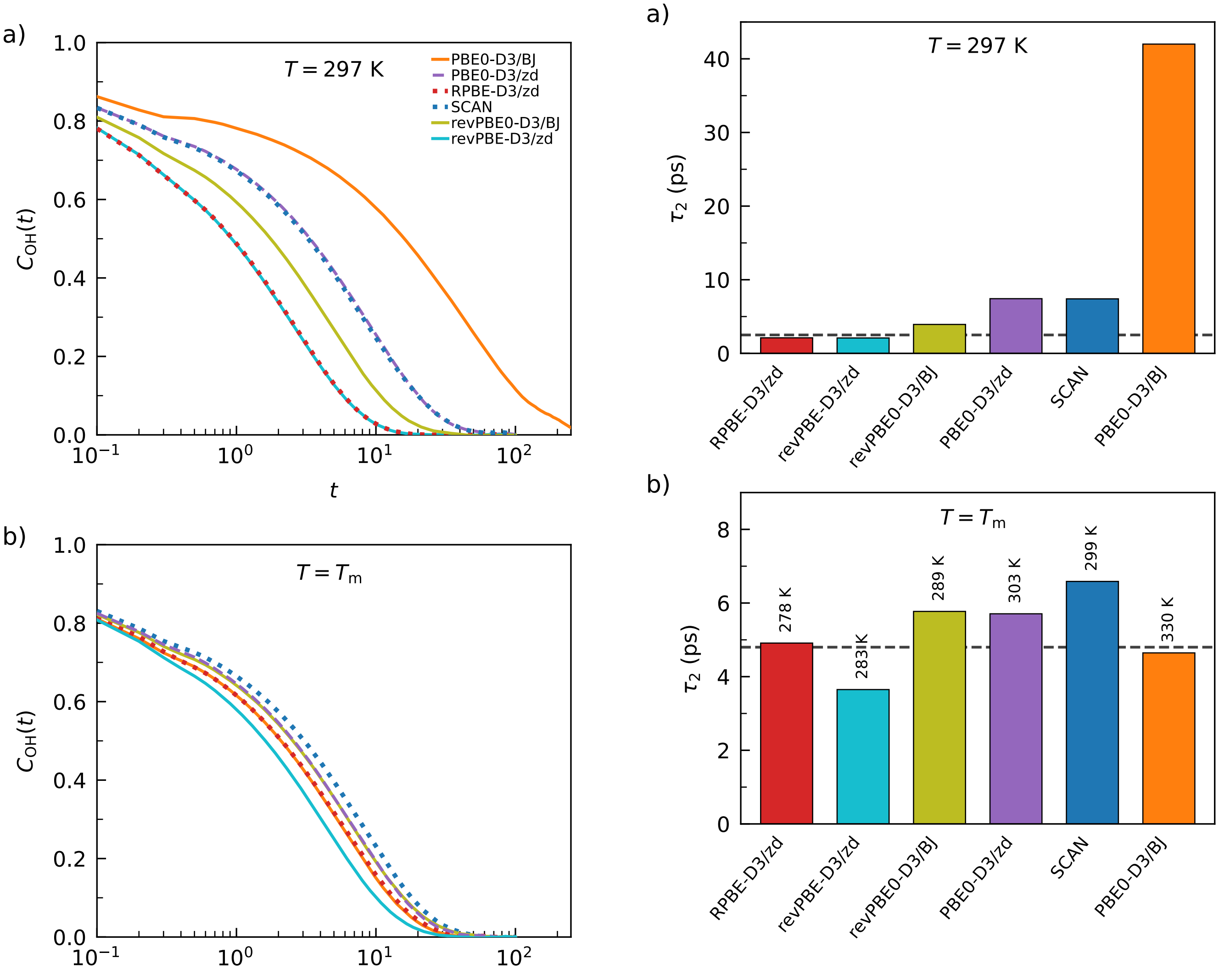

FIG. 5. Normalized OH bond rotational correlation function, $C_{\mathrm{OH}}(t)$, for the different functionals. Panel a) shows the correlation functions at $T = 297$ K, while panel b) shows the corresponding results at each model's melting temperature. Different colors and line styles distinguish the functionals.

FIG. 6. Comparison of the second-rank relaxation time $\tau_2$ values obtained with different density functionals at two thermodynamic conditions: a) $T = 297$ K, and b) $T = T_{\mathrm{m}}$ K. The dashed horizontal lines indicate experimental estimates at similar conditions, i.e., at 274.15 [19] for $T = T_{\mathrm{m}}$, and 298.15 K [29] to be compared with 297 K. Melting temperatures of the corresponding functionals used in panel b) are indicated above the bars and taken from Ref. [6].

In Fig. 8a), we report the diffusion coefficient $D$ along different isotherms as a function of pressure, together with experimental data. Given the wide range of temperatures and pressures explored, the overall agreement is good. Deviations are most noticeable at 250 K near ambient pressure, although the agreement of this isotherm improves at higher pressures. The simulations reproduce the characteristic maximum in $D$, as well as its shift with pressure across isotherms, in qualitative agreement with experiment. Importantly, the model also captures the correct order of magnitude of $D$ over the full thermodynamic range. Specifically, the agreement for the 270 K isotherm is quantitatively very good.

The reported values of $D$ were corrected for finite-size effects using Eq. (11), employing the viscosity obtained from the same simulations. The corresponding values of $\eta$ are shown in Fig. 8b). Similar to the behavior observed for $D$, the agreement with experiment is generally good, with noticeable deviations only at low temperature near ambient pressure. For the remaining isotherms, the agreement is close to quantitative.

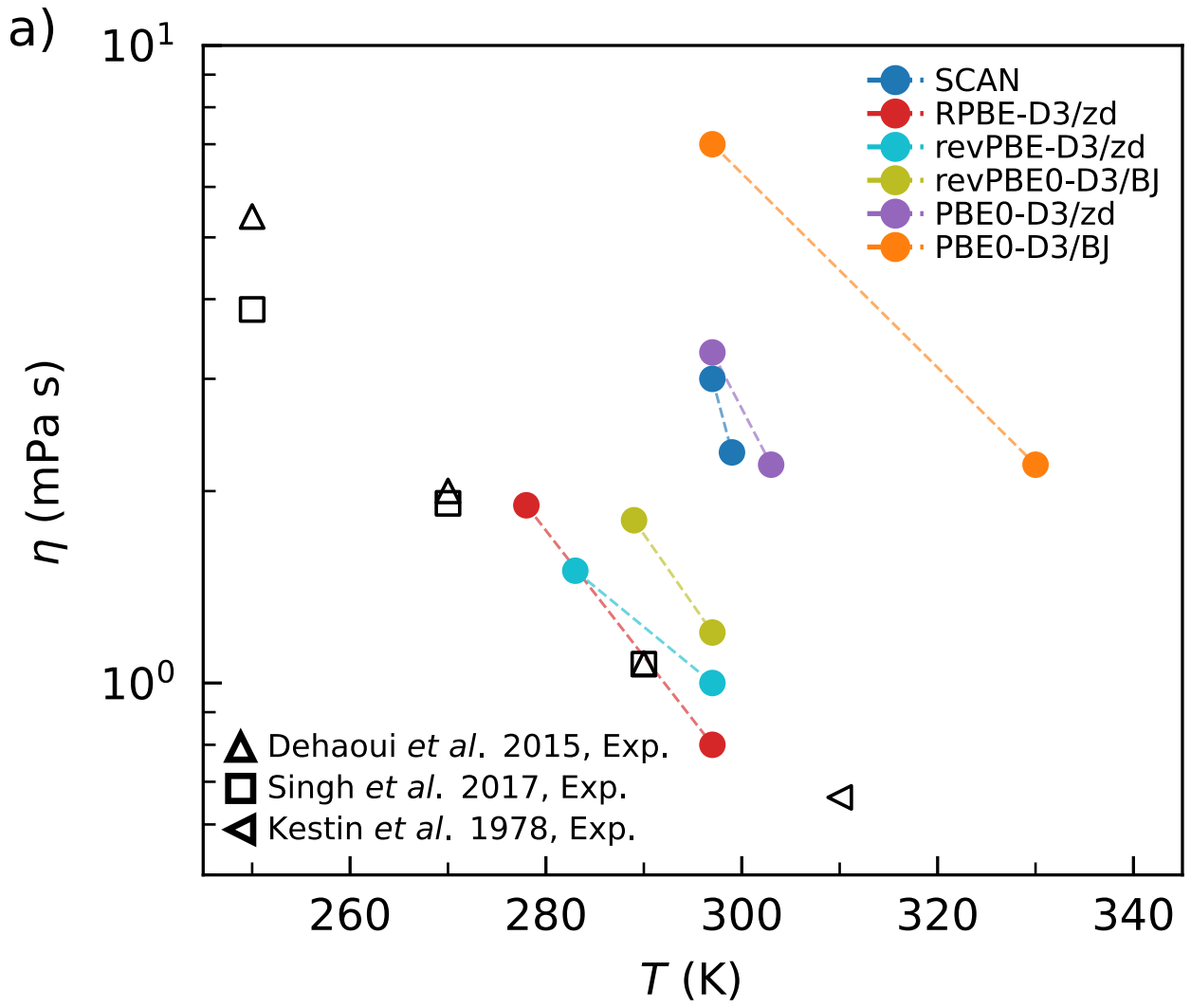

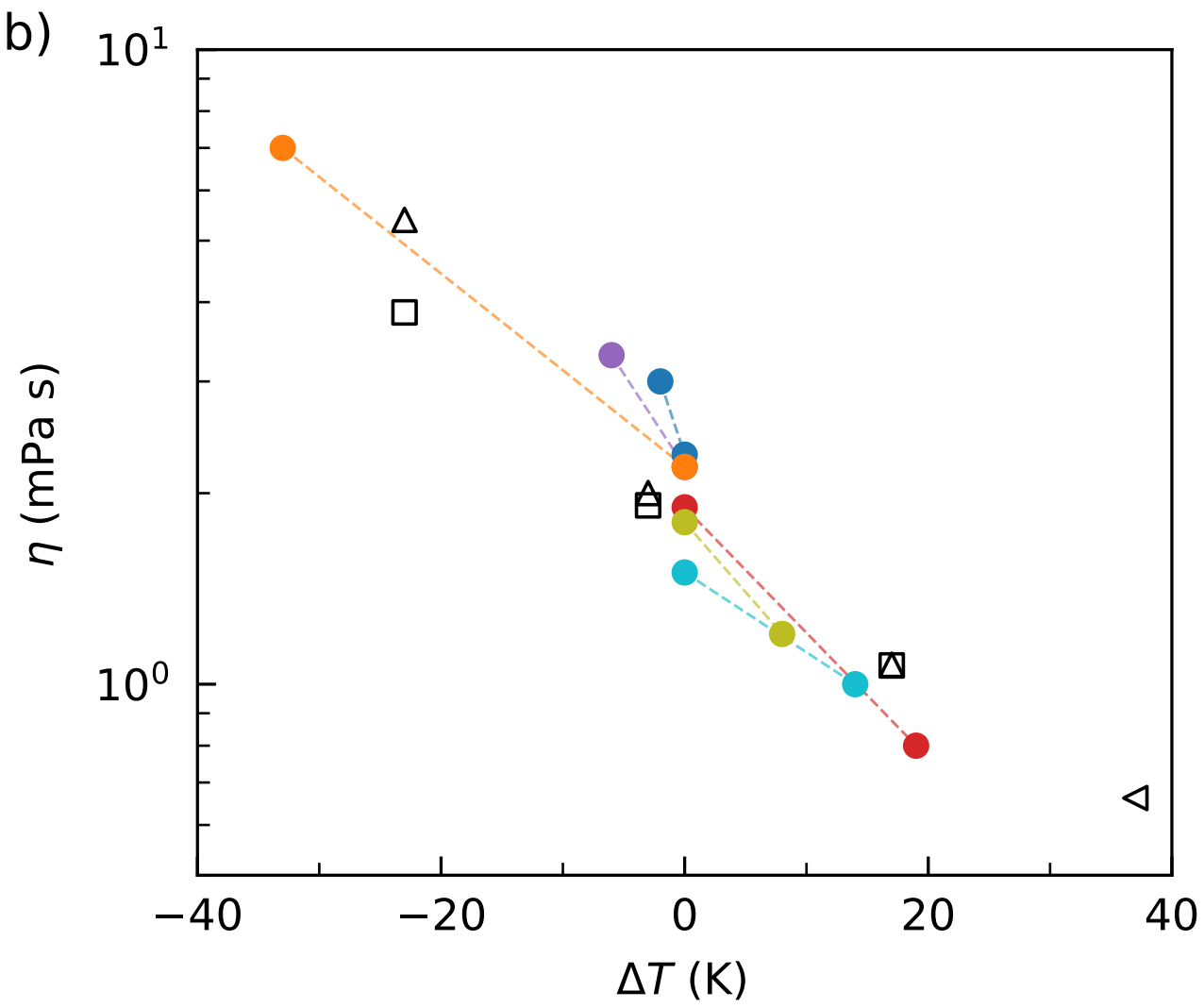

FIG. 7. a) Shear viscosity $\eta$ of water as a function of temperature $T$ from experiments and simulations based on different density functionals. Experimental data around ambient pressure, $|p| \leq 50$ MPa, are shown in black, where upward triangles, squares, and left-pointing triangles correspond to Refs. [16], [20], and [17], respectively. Simulation results are shown with filled circles; colors identify the underlying density functional. b) Same $\eta$ values as in panel a), plotted as a function of the supercooling $\Delta T = T - T_{\mathrm{m}}$, where $T_{\mathrm{m}}$ is the melting temperature of the corresponding functional.

In panel c) of Fig. 8, we report the rotational relaxation time $\tau_2$ for the same set of state points. Interestingly, at high pressures $\tau_2$ remains nearly constant, whereas $D$ decreases and $\eta$ increases. At lower pressures, $\tau_2$ increases, consistent with the behavior of $\eta$ along low-temperature isotherms, but not along higher-temperature ones, where $D$ still decreases. The interplay between these kinetic observables has been discussed extensively for other liquids and water models [12, 13, 42]. To further assess the rotational dynamics, we compute $\tau_2$ at ambient pressure over a range of temperatures spanning from near the melting point to higher temperatures, and compare with three experimental data points. As shown in Fig. 8d), the agreement is very good overall, with deviations becoming pronounced as the temperature approaches $T_{\mathrm{m}}$.

The pressure-dependent data provide a more stringent test than the ambient-pressure comparison, because the model must reproduce not only the magnitude of the transport coefficients but also their variation across different regions of the liquid-state phase diagram. The qualitative agreement with experiment is therefore encouraging. In particular, the simulations capture the non-monotonic pressure dependence of $D$, including the presence of a maximum and its shift with temperature. This behavior is a characteristic anomaly of liquid water and is closely connected to the competition between the open tetrahedral local structure and denser liquid arrangements [3].

The comparison between $D$, $\eta$, and $\tau_2$ also shows that the different dynamical observables do not respond identically to compression. While $D$ and $\eta$ are broadly anticorrelated, as expected from hydrodynamic arguments, the rotational relaxation time displays a weaker pressure dependence at high pressures. This suggests that translational mobility, viscous relaxation, and local reorientation are affected slightly differently by the structural changes induced by compression. Therefore, agreement with experiment for one transport property does not automatically imply agreement for all dynamical observables, making the simultaneous comparison of $D$, $\eta$, and $\tau_2$ a useful validation of the RPBE-D3/zd functional.

## IV. DATA

In this section, we tabulate the numerical values shown in the figures of this work. We first summarize the diffusion coefficients and viscosities obtained for the different density functionals in Tab. II. This comparison is complemented by an additional table containing the corresponding rotational relaxation times in Tab. III. We then report the results obtained for RPBE-D3/zd with the BPNNP over the broader range of thermodynamic conditions considered in this study in Tabs. IV and V.

TABLE II. Diffusion coefficients $D$ and shear viscosities $\eta$ of liquid water obtained with different density functionals using MACE models. Entries without a reported viscosity correspond to NVE simulations of a few ns, for which the pressure may deviate from ambient conditions, although it remains within ±350 bar. Viscosity values are reported for the 600 ps NVT simulations. For these NVT state points, the corresponding diffusion coefficients were corrected for finite-size effects using Eq. (11).

| Functional | $T$ (K) | $D$ (nm$^2$ ns$^{-1}$) | $\eta$ (mPa s) |
|---|---|---|---|
| SCAN | 280.0 | 0.177 | – |
| SCAN | 307.0 | 0.894 | – |
| SCAN | 299.0 | 1.047 | 2.3 |
| RPBE-D3/zd | 269.5 | 0.689 | – |
| RPBE-D3/zd | 296.0 | 2.065 | – |
| RPBE-D3/zd | 315.0 | 3.661 | – |
| RPBE-D3/zd | 278.0 | 1.484 | 1.9 |
| RPBE-D3/zd | 297.0 | 2.768 | 0.8 |
| revPBE-D3/zd | 273.0 | 0.771 | – |
| revPBE-D3/zd | 291.5 | 1.998 | – |
| revPBE-D3/zd | 313.0 | 3.514 | – |
| revPBE-D3/zd | 283.0 | 1.742 | 1.5 |
| revPBE-D3/zd | 297.0 | 2.688 | 1.2 |
| revPBE0-D3/BJ | 272.0 | 0.343 | – |
| revPBE0-D3/BJ | 292.0 | 0.952 | – |
| revPBE0-D3/BJ | 317.0 | 2.390 | – |
| revPBE0-D3/BJ | 289.0 | 1.200 | 1.8 |
| revPBE0-D3/BJ | 297.0 | 1.593 | 1.2 |
| PBE0-D3/zd | 291.0 | 0.455 | – |
| PBE0-D3/zd | 312.0 | 0.969 | – |
| PBE0-D3/zd | 326.0 | 1.551 | – |
| PBE0-D3/zd | 297.0 | 0.805 | 3.3 |
| PBE0-D3/zd | 303.0 | 1.029 | 2.2 |
| PBE0-D3/BJ | 297.0 | 0.411 | 7.0 |
| PBE0-D3/BJ | 330.0 | 1.371 | 2.2 |

TABLE III. Second-order rotational relaxation times $\tau_2$ obtained with different density functionals for water using MACE models at $T = 297$ K and at melting temperature $T_m$ of each functional, see Tab. I for the values of $T_m$.

| Functional | $\tau_2^{297\,\mathrm{K}}$ (ps) | $\tau_2^{T_m}$ (ps) |
|---|---|---|
| RPBE-D3/zd | 2.112 | 4.912 |
| revPBE-D3/zd | 2.090 | 3.649 |
| revPBE0-D3/BJ | 3.930 | 5.771 |
| PBE0-D3/zd | 7.429 | 5.707 |
| SCAN | 7.403 | 6.586 |
| PBE0-D3/BJ | 42.000 | 4.645 |

TABLE IV. Diffusion coefficient $D$, shear viscosity $\eta$, and second-order rotational relaxation time $\tau_2$ for RPBE-D3/zd water along different isotherms.

| $T$ (K) | $\rho$ (kg m$^{-3}$) | $p$ (MPa) | $D$ (nm$^2$ ns$^{-1}$) | $\eta$ (mPa s) | $\tau_2$ (ps) |
|---|---|---|---|---|---|
| 310 | 800 | -160.933 | 3.647 | 0.550 | 1.655 |
| 310 | 900 | 33.147 | 3.835 | 0.578 | 1.350 |
| 310 | 1000 | 360.315 | 3.196 | 0.654 | 1.319 |
| 310 | 1100 | 969.075 | 2.519 | 0.829 | 1.289 |
| 290 | 800 | -182.774 | 1.761 | 1.023 | 4.166 |
| 290 | 900 | 20.493 | 2.176 | 1.038 | 2.876 |
| 290 | 1000 | 327.979 | 2.109 | 1.023 | 1.912 |
| 290 | 1100 | 900.252 | 1.555 | 1.527 | 2.131 |
| 270 | 800 | -196.231 | 0.481 | 5.195 | 15.890 |
| 270 | 900 | 21.374 | 0.854 | 2.481 | 6.426 |
| 270 | 1000 | 308.803 | 1.184 | 1.538 | 3.799 |
| 270 | 1100 | 840.430 | 0.826 | 2.336 | 4.033 |
| 250 | 900 | 58.682 | 0.233 | 10.872 | 27.296 |
| 250 | 1000 | 299.287 | 0.510 | 3.557 | 8.859 |
| 250 | 1100 | 778.479 | 0.367 | 6.041 | 9.506 |

## V. CONCLUSIONS

We have compared the dynamical properties of liquid water predicted by several density functionals using machine-learning interatomic potentials. For this purpose, we trained MACE models on available datasets [6] corresponding to SCAN, RPBE-D3/zd, revPBE-D3/zd, revPBE0-D3/BJ, PBE0-D3/zd, and PBE0-D3/BJ, and computed translational, rotational, and viscous properties. The analysis combined time-correlation functions, which probe the microscopic relaxation mechanisms, with long-time kinetic coefficients, which provide direct comparison with experimental measurements.

The comparison among functionals shows that the predicted dynamics vary substantially, as previously observed for the density [6]. At a fixed temperature, part of the spread reflects the fact that each functional presents a different stability relative to its own melting point. This is suggested by the improved agreement obtained when diffusion coefficients, rotational relaxation times, and viscosities are considered as functions of the temperature relative to the melting point. Nevertheless, the rescaling by $T_\mathrm{m}$ does not provide a universal correction. Differences remain even at comparable thermodynamic stability with respect to ice, i.e., at the melting point, indicating that the functionals also differ in their underlying description of the liquid-state free-energy landscape and microscopic relaxation mechanisms.

Among the functionals considered, RPBE-D3/zd provides the best overall agreement with experiment. It performs well for translational diffusion, rotational relaxation, and shear viscosity, closely followed by revPBE-D3/zd, which also gives good diffusion coefficients, but its agreement is slightly less consistent when rotational dynamics are considered. SCAN and PBE0-

TABLE V. Second-order rotational relaxation time $\tau_2$ for RPBE-D3/zd water along the ambient-pressure isobar as a function of temperature.

| $T$ (K) | $\tau_2$ (ps) |
|---|---|
| 273 | 9.213 |
| 280 | 7.335 |
| 283 | 4.525 |
| 287 | 3.673 |
| 295 | 2.282 |
| 310 | 1.529 |
| 325 | 0.963 |
| 340 | 0.693 |
| 350 | 0.591 |

D3/zd predict dynamics that are too slow compared with real water, while PBE0-D3/BJ appears strongly shifted along the phase diagram. revPBE0-D3/BJ shows intermediate behavior and is the most accurate of the hybrid functionals considered here.

The broader comparison of RPBE-D3/zd with experiment over a wide range of temperatures, densities, and pressures further supports this conclusion. The model reproduces the correct order of magnitude of the diffusion coefficient and captures its non-monotonic pressure dependence, including the presence of a maximum and its shift with temperature. The viscosities are also in good agreement with experiment over most of the thermodynamic range. The largest deviations occur at low temperature near ambient pressure, where the dynamics become particularly sensitive to the proximity of the melting region. The rotational relaxation time $\tau_2$ is also well described overall, especially along the ambient-pressure isobar, although the comparison across pressure shows that translational, viscous, and rotational relaxation do not respond identically to compression.

Overall, these results show that dynamical observables provide a stringent and complementary benchmark for first-principles water models. A better description of thermodynamic properties, e.g., density, is not sufficient to guarantee a better description of transport. Within the set of functionals studied here, RPBE-D3/zd offers the most balanced description of liquid-water dynamics and therefore remains a strong candidate for simulations of aqueous systems where transport properties are relevant.

## AUTHOR INFORMATION

### Corresponding Author

Pablo Montero de Hijes – Faculty of Physics, University of Vienna, A-1090 Vienna, Austria; Email: pablo.montero.de.hijes@univie.ac.at

### Authors

Leon Neubeck – Faculty of Physics, University of Vienna, A-1090 Vienna, Austria
Georg Kresse – Faculty of Physics, University of Vienna, A-1090 Vienna, Austria; VASP Software GmbH, Berggasse 21, A-1090 Vienna, Austria
Christoph Dellago – Faculty of Physics, University of Vienna, A-1090 Vienna, Austria

### Notes

The authors declare no competing financial interests.

## ACKNOWLEDGMENTS

This research was funded in part by the Austrian Science Fund (FWF) through the SFB TACO 10.55776/F8100 and in part by 10.55776/COE5 (Cluster of Excellence MECS). For open access purposes, the author has applied a CC BY public copyright license to any author-accepted manuscript version arising from this submission. Computer resources and technical assistance were provided by the Austrian Scientific Computing (ASC).

## AUTHOR DECLARATIONS

### Conflict of Interest

The authors have no conflicts to disclose

### Data availability

The data that support the findings of this paper are available upon request.

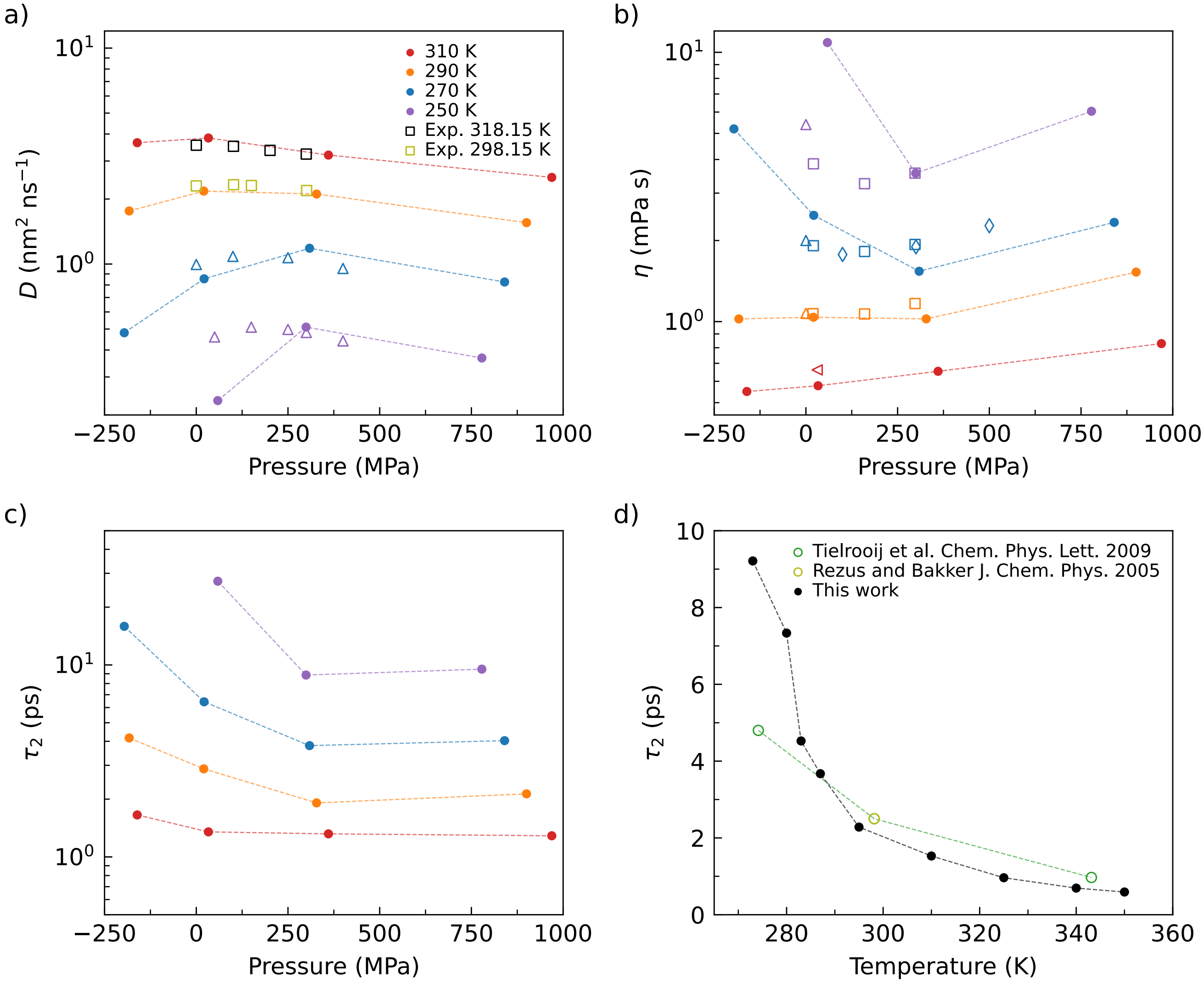

FIG. 8. a) Self-diffusion coefficient $D$, b) viscosity $\eta$, and c) second-rank rotational relaxation time $\tau_2$ of water as functions of pressure along different isotherms, comparing experimental data with *ab initio* machine-learning simulations based on RPBE-D3/zd. Results from this work are shown as filled symbols connected by dashed lines, whereas literature data are represented by open symbols without connecting lines. In panel a), experimental data are shown as triangles [43] and squares [44]. In panel b), upward triangles, left-pointing triangles, squares, and diamonds correspond to Refs. [16], [17], [20], and [45], respectively. Although colors denote temperature, experimental temperatures do not exactly coincide with the simulated isotherms, but typically differing by less than 3 K. An exception is the dataset for $D$ from Ref. [44], shown in darker shades of orange and red, corresponding to 298.15 K and 318.15 K, respectively. In panel d), $\tau_2$ is shown as a function of temperature and compared with the experimental results of Refs. [19] and [29] at ambient pressure.